\documentclass[prl,superscriptaddress,twocolumn,showpacs]{revtex4}

\usepackage{graphicx}
\usepackage{dcolumn}
\usepackage{bm}

\usepackage{color}  

\newcommand{\cut}[1] {\textcolor{green}{[----]}}

\begin{document}

\preprint{YBCO ortho-II/LaTeX V.1}

\title{X-ray diffraction observation of charge density wave order at zero magnetic field in ortho-II YBa$_2$Cu$_3$O$_{6.54}$}

\author{E. Blackburn}
\affiliation{School of Physics and Astronomy, University of Birmingham, Birmingham B15 2TT, United Kingdom.}
\author{J. Chang}
\email{johan.chang@epfl.ch}
\affiliation{Institut\,de\,la\,mati\`ere\,complexe,\,Ecole\,Polytechnique\,F\'ederale\,de\,Lausanne (EPFL),\,CH-1015\,Lausanne,\,Switzerland.}
\affiliation{Swiss Light Source, Paul Scherrer Institut, CH-5232 Villigen PSI, Switzerland}

\author{M. H\"ucker}
\affiliation{Condensed Matter Physics \& Materials Science Dept., Brookhaven National Lab., Upton, NY 11973, USA.}

\author{A. T. Holmes}
\affiliation{School of Physics and Astronomy, University of Birmingham, Birmingham B15 2TT, United Kingdom.}

\author{N. B. Christensen}
\affiliation{Department of Physics, Technical University of Denmark, DK-2800 Kongens Lyngby, Denmark.}

\author{Ruixing Liang}
\affiliation{Department of Physics $\&$ Astronomy, University of British Columbia, Vancouver, Canada.}
\affiliation{Canadian Institute for Advanced Research, Toronto, Canada.}

\author{D. A. Bonn}
\affiliation{Department of Physics $\&$ Astronomy, University of British Columbia, Vancouver, Canada.}
\affiliation{Canadian Institute for Advanced Research, Toronto, Canada.}

\author{W. N. Hardy}
\affiliation{Department of Physics $\&$ Astronomy, University of British Columbia, Vancouver, Canada.}
\affiliation{Canadian Institute for Advanced Research, Toronto, Canada.}

\author{U. R\"utt}
\affiliation{Deutsches Elektronen-Synchrotron DESY, 22603 Hamburg, Germany.}

\author{O. Gutowski}
\affiliation{Deutsches Elektronen-Synchrotron DESY, 22603 Hamburg, Germany.}

\author{M.~v.~Zimmermann}
\affiliation{Deutsches Elektronen-Synchrotron DESY, 22603 Hamburg, Germany.}

\author{E. M. Forgan}
\affiliation{School of Physics and Astronomy, University of Birmingham, Birmingham B15 2TT, United Kingdom.}

\author{S. M. Hayden}
\affiliation{H. H. Wills Physics Laboratory, University of Bristol, Bristol, BS8 1TL, United Kingdom.}
\date{\today}
\begin{abstract}
X-ray diffraction measurements show that the high-temperature superconductor  YBa$_2$Cu$_3$O$_{6.54}$, with ortho-II oxygen order, 
has charge density wave order (CDW) in the absence of an applied magnetic field.  The dominant wavevector of the CDW is 
$\mathbf{q}_{\mathrm{CDW}} = (0, 0.328(2), 0.5)$, with the in-plane component  parallel to the $\mathbf{b}$-axis (chain direction). It has a 
similar incommensurability to that observed in ortho-VIII and ortho-III samples, which have different dopings and oxygen orderings.  
Our results for ortho-II contrast with recent high-field NMR measurements, which suggest a commensurate wavevector along the $\mathbf{a}$-axis. 
 We discuss the relationship between spin and charge correlations in YBa$_2$Cu$_3$O$_{y}$, and recent high-field quantum oscillation, NMR and ultrasound experiments.
\end{abstract}

\maketitle


Charge density waves (CDWs) have recently been observed in the high temperature superconductors (HTS) YBa$_2$Cu$_3$O$_{y}$ (YBCO) 
and (Y/Nd)Ba$_2$Cu$_3$O$_{y}$ \cite{Ghiringhelli,Chang,Achkar}.  The CDW, which competes with HTS, develops in a region inside the
 celebrated pseudogap phase, where a number of other probes, including high-field NMR \cite{Wu}, the Kerr effect \cite{Xia}, and
 the Hall effect \cite{LeBoeufPRB}, show signatures of electronic ordering. This last effect is smoothly connected to the
 low-temperature high-field quantum oscillations (QO) \cite{Doiron} that demonstrated the existence of small Fermi surface (FS) pockets.


The existence of ground states with competing order is central to many theories of HTS. A widely discussed example is ``stripe order'', that is a state with coexisting charge and spin order \cite{Kivelson}. Stripe order is observed in some HTS and related compounds, such as La$_{2-x}$Ba$_{x}$CuO$_4$ \cite{Hucker} and La$_{1.6-x}$Nd$_{0.4}$Sr$_{x}$CuO$_4$ \cite{Tranquada}.  It is important to establish whether the tendency towards stripes is a generic property of the cuprates and whether the spin and charge correlations are always related.

YBa$_2$Cu$_3$O$_{y}$ differs from, e.g.~La$_{2-x}$(Ba,Sr)$_{x}$CuO$_4$, in that it contains bilayers of CuO$_2$ planes, separated by
 layers containing a certain fraction (depending on $y$) of Cu-O chains. The oxygen-filled chains, which run along the orthorhombic crystal
 {\bf b}-direction tend to order and are labelled ortho-$N$, depending on the repeat length ($Na$) of the ordering of the chains along {\bf a} \cite{Fontaine,Beyers,Zimmermann}.
A major gap in the CDW picture to date was the failure to observe a CDW in the ortho-II state (the most highly ordered, having alternating full and empty Cu-O chains).  This was surprising because many 
studies of this composition had suggested that such order is present, at least in high field \cite{Wu, LeBoeufPRB, LeBoeuf, Laliberte, Sebastian, Vignolle}. Most recently, ultrasound measurements 
have indicated a 2-{\bf q} state in high magnetic fields \cite{LeBoeufNP}.

In this Letter, we report the observation of a CDW in an ortho-II sample of YBa$_2$Cu$_3$O$_{6.54}$ in zero 
magnetic field with dominant wavevector $\mathbf{q}_{\mathrm{CDW}} = (0, 0.328(2), 0.5)$.  
This contrasts with both lower and higher dopings, where two wavevectors are observed corresponding to two modulations of similar amplitude \cite{Ghiringhelli,Chang,Achkar}.
Both the wavevector magnitude and direction of $\mathbf{q}_{\mathrm{CDW}}$ differ from those inferred from NMR measurements
in high field \cite{Wu}. The propagation vectors of the charge and spin correlations in ortho-II
 YBCO do not appear to follow the simple relationship
 $ \boldsymbol{\delta}_{\mathrm{charge}}= 2\boldsymbol{\delta}_{\mathrm{spin}}$ observed
 in stripe systems \cite{Hucker,Tranquada}.  We examine the doping dependence of the
 CDW order in high quality ortho-II, -III and -VIII samples
(hereafter denoted o-II, o-III and o-VIII).


\begin{figure*}[th]
\begin{center}
\includegraphics[width=0.8\textwidth]{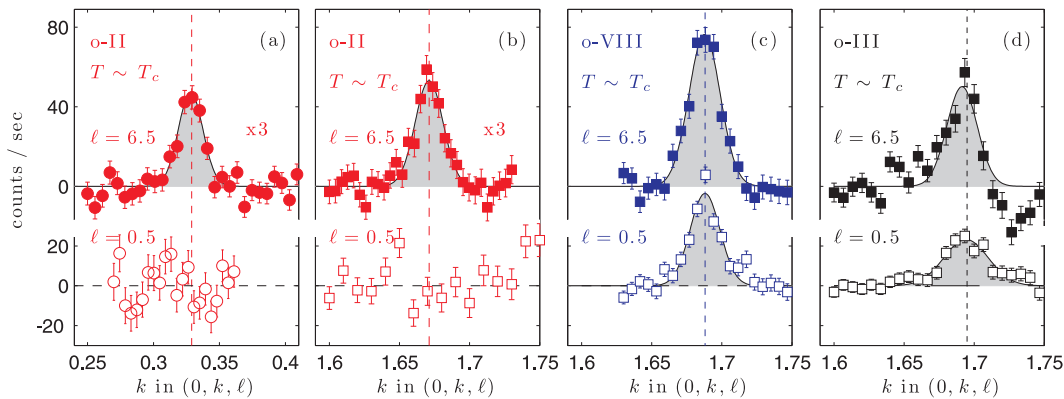}
\end{center}
\caption{
$k$- scans through the CDW wavevector positions $(0,\delta_2,\ell)$ and  $(0,2-\delta_2,\ell)$ in
YBCO o-II, -VIII, and -III at $T \sim T_{\mathrm{c}}$. (a, b) For o-II,  $\delta_2=0.328(2)$ incommensurate peaks are only 
observed at $\ell=6.5$, suggesting a strong CDW displacement along $z$. (c, d) For o-VIII and o-III, peaks are observed at $\ell$=0.5 
and 6.5 with  $\delta_2$ = 0.314(2) and 0.308(2) respectively. Linear backgrounds have been subtracted in panels (a) and (b).  To avoid 
contamination from weakly $T$-dependent spurious peaks in (c) and the o-III superlattice peak from the minority domain in (d), we 
plot $I(T_{\mathrm{c}}) - I(140~\mathrm{K})$ corrected for a
sloping background as in (a, b). Intensities in (a, b) have been multiplied by a factor 3 to compensate for the smaller  o-II sample size.  }
\label{fig:fig1}
\end{figure*}

We carried out high energy (100 keV) X-ray diffraction experiments on  three $\sim 99\%$  detwinned YBa$_2$Cu$_3$O$_y$ single crystals (sample characteristics are given in Table.~1).
These samples have orthorhombic crystal structures ($a\approx3.82$ (ignoring the chain-ordering superlattices), $b\approx3.87$ and $c\approx11.7$~\AA).
The o-II sample shows more perfect oxygen-chain order than the other compositions.  The coherence lengths are $\xi_a \sim 110$ \AA, $\xi_b > 150$ \AA,  and $\xi_c=55$ \AA\
for o-II; $\xi_a \sim 20$~\AA\  and $\xi_c<10$~\AA\ for o-III and $\xi_a \sim 25$~\AA\  and $\xi_c<10$~\AA\ for o-VIII.  Our samples are highly uniform, 
strongly indicating that the CDW occupies the entire sample volume and not a fraction.

Samples were mounted in a closed-cycle cryostat on a 4-circle diffractometer on beamline BW5
 at the DORIS storage ring (DESY). This allowed access to a wide range of
 reciprocal space $(h, k, \ell)$  expressed in units of $(2\pi/a, 2\pi/b, 2\pi/c)$, at temperatures down to 6~K.

 \begin{table}[htb]
 \begin{center}
\begin{ruledtabular}
 \begin{tabular}{ccccccc}
 $y$ in&Oxygen & Doping & $T_{\mathrm{c}}$  & $T_{\mathrm{CDW}}$  & $\delta_1 ({\bf a})$  & $\delta_2$ ({\bf b}) \\
 YBCO&ordering & level $p$ & (K) & (K)& (r.l.u)&(r.l.u) \\\hline
 6.54&o-II & 0.104 & 58 & 155(10) & 0.320(2) & 0.328(2)  \\
 6.67&o-VIII & 0.123 & 67 & 140(10) & 0.305(2) & 0.314(2)  \\
 6.75&o-III & 0.132 & 74 & 140(10) & 0.30 \cite{Achkar} & 0.308(2)  \\ 
 \end{tabular}
\end{ruledtabular}
 \caption{Characteristics of the YBa$_2$Cu$_3$O$_{y}$ samples studied.
The superconducting $T_c$ was determined from the 1~Oe field-cooled magnetization
and doping was evaluated from Ref.~\onlinecite{Ruixing}.
$T_{\mathrm{CDW}}$, $\delta_1$ and $\delta_2$ are derived from
our high energy X-ray experiments -- see also the note added at the end of this Letter --
 (quoted uncertainties in the values of $\delta$ are dominated by minor crystal alignment errors) and from Ref. \onlinecite{Achkar}.}
\end{center}
 \label{tab:tab1}
 \end{table}

A CDW modulation with characteristic wavevector $\mathbf{q}_{\mathrm{CDW}}$ gives
 rise to satellites around reciprocal lattice points at positions
 $\mathbf{Q}=\boldsymbol{\tau} \pm \mathbf{q}_{\mathrm{CDW}}$. High energy X-ray
diffraction is sensitive to the atomic displacements
parallel to the scattering vector $\mathbf{Q}$ \cite{Chang}. Our previous measurements \cite{Chang} on o-VIII showed $\mathbf{q}_{\mathrm{CDW}}$=$(\delta_1, 0, 0.5)$ and $(0, \delta_2, 0.5)$ (see Table~1).
Fig.~1(a,b) shows $k$-scans performed on YBCO o-II through
the positions $(0,\delta, \ell)$ and $(0, 2-\delta, \ell)$,
with $\ell =$ 0.5, 6.5 and $\delta \sim 0.3$.
No CDW peaks were found at $\ell=0.5$, but well-defined
peaks were observed at $\ell=6.5$. These peaks constitute the
 first direct X-ray evidence for CDW order in YBCO o-II  and
indicate that the displacements are mainly polarised along {\bf c}.
For comparison, $k$-scans  through the positions $(0, 2-\delta, \ell)$ in
 o-VIII and o-III are shown in Fig.~1(c, d) for $\ell = 0.5$ and 6.5;
again, for modulations along $k$, the signal at $\ell = 6.5$ is stronger.

 In YBCO o-VIII and -III, a lattice modulation is found along both
the {\bf a}- and {\bf b}-axis directions \cite{Ghiringhelli,Chang,Achkar}.
For o-II, we have searched for a CDW along the {\bf a}-axis,
concentrating on positions where a signal was observed in o-VIII.
Wavevectors ($| n-\delta |, 0, \ell$) with $n=2,4$ for $\ell=0.5$ and
$n=0, 2, 4$ for $\ell=6.5$ were measured and no signal from a lattice modulation
was found above the noise level (see Fig. 2(a, b)). However, see also note added in 
proof at the end.

Even though the CDW structures in YBCO o-II, -III and -VIII
are different, the signals from the {\bf b}-axis modulation
 have very similar dependence on $\ell$, $T$ and magnetic field.
Fig.~3(a) shows the $\ell$-dependence near $\ell = 6.5$ at $T\sim T_{\mathrm{c}}$.
All three compounds have a broad peak at $\ell \approx 6.5$, with a width corresponding to
a correlation length of $\xi_c\lesssim 10$~\AA.  
The $T$-dependence of several reflections is
plotted in Fig. 3 (c). As was previously shown for YBCO
o-VIII and o-III~\cite{Chang,Ghiringhelli,Achkar},
the intensity grows below an onset temperature $T_{\mathrm{CDW}}$ $(\sim140~\mathrm{K})$
down to the respective $T_c$'s, below which a partial suppression
takes place. Here we show a similar behaviour in o-II, but with a slightly higher onset temperature.
The application of a magnetic field
enhances the low-$T$ intensity in a similar fashion to that
reported in YBCO o-VIII~\cite{Chang}.
It is interesting to note [see Fig.~\ref{fig:fig3}(b)] that the width of the Bragg peak becomes smaller in 
a  field with 11.5~T along {\bf c}, corresponding to an increase of correlation length $\xi_b$ from 48$\pm$9~\AA\ to 63$\pm$7~\AA. 
The presence of a magnetic field suppresses superconductivity and the resurgent competing CDW increases its correlation length.

Modelling of our {\bf b}-direction data for all
three dopings suggests that the displacement pattern involves {\bf c} axis
displacements of the bilayer oxygens, similar to those proposed
for the soft phonon in YBa$_2$Cu$_3$O$_{7}$~\cite{Raichle}. 
There are small differences between the values of the modulation
periods for the {\bf a} and {\bf b} directions for all three compounds (Table 1) and in the
patterns of atomic displacements, showing that the influence
of the chains on the planes is also noticeable in o-VIII and o-III.
If the two distortions develop independently, one would expect a different $T_{\mathrm{CDW}}$ 
for modulations along each direction, with the postulated 2-{\bf q} state forming at lower temperatures. 
 To date, we have no evidence for this, although observations in o-II most clearly indicate a difference between the 
CDW order in the {\bf a}- and {\bf b}- directions.  The in-plane electronic anisotropy in YBCO arises from the chains, 
thus they must ultimately be responsible for this difference between the 
{\bf a}- and {\bf b}- directions. An obvious mechanism is through the chain Fermi surface~\cite{CY} with spanning 
vectors along {\bf b}$^*$, which might encourage CDW formation. We note that STM observations on the chain-surface 
of optimally-doped YBCO~\cite{DerroPRL2002}, show such behaviour, with a $\delta_2 \sim 0.3$. Furthermore, the alternating 
filled and empty chains create an additional potential which would fold the Fermi surface along {\bf a}$^*$ and thereby 
change the band structure~\cite{CY}.

\begin{figure}[t]
\begin{center}
\includegraphics[width=0.85\linewidth]{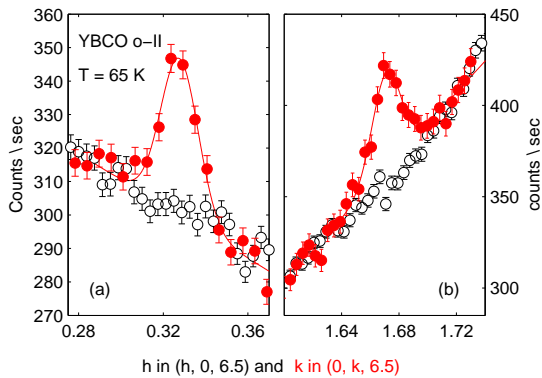}
\end{center}
\caption{(a)-(b) $h-$ and $k-$scans, taken on o-II, through
($\mid n-\delta_1 \mid$, 0, 6.5) and (0, $\mid n-\delta_2\mid$, 6.5) with $n=0,2$
 and $T \sim T_{\mathrm{c}}$.
 The $k$-scans (red circles), showing lattice modulation peaks
at (0, $\delta_2$, 6.5) and (0, 2-$\delta_2$, 6.5), are the same as
displayed in  Fig.~\ref{fig:fig1}(a,b). Equivalent measurements in the ($h$, 0, $\ell$)-plane
(black circles) reveal no evidence for a lattice modulation at $\delta_1 \sim \delta_2$. 
Notice that the lattice modulation peaks are two orders of magnitude weaker 
than the reflections from the ortho-II structure. 
}
\label{fig:fig2}
\end{figure}

\begin{figure}[thb]
\begin{center}
\includegraphics[width=0.98\linewidth,clip=true]{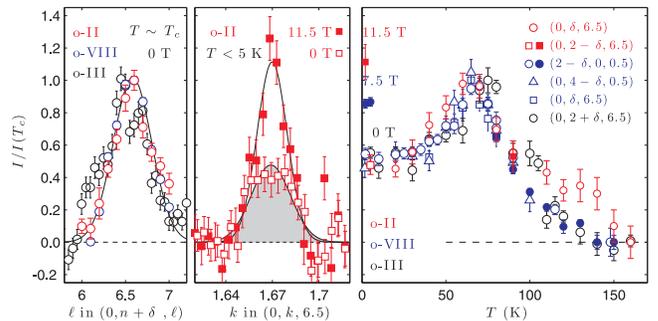}
\end{center}
\caption{Out-of-plane momentum $\ell$-, field- and temperature-dependence
of the CDW modulation peaks found in YBCO o-II (red), o-VIII (blue) and o-III (black). All intensities have been background-subtracted and normalized to $I(T_{\mathrm{c}})$ in zero field.
(a) $\ell$-dependence of the peak height of $k$-scans through $(0, n+\delta_2, \ell)$
 with $n = 0$ for o-II and o-VIII, and $n = 2$ for o-III. All compounds
 show a broad peak centered at $\ell \sim 6.5$ and a {\bf c}-axis correlation length
 $\xi_c$  comparable to that previously
 reported \cite{Chang} in o-VIII at $T$ = 2~K for $\ell = 0.5$.
(b) Measurement in a separate cryostat of the effect on CDW intensity in o-II of a magnetic field applied with a component 11.5 T along the {\bf c}-axis of the crystal.
(c) Temperature dependence of peak intensities, measured at the wave vectors indicated. Filled symbols are data taken in a magnetic field. }
\label{fig:fig3}
\end{figure}

\begin{figure}[thb]
\begin{center}
\includegraphics[width=0.9\linewidth,clip=true]{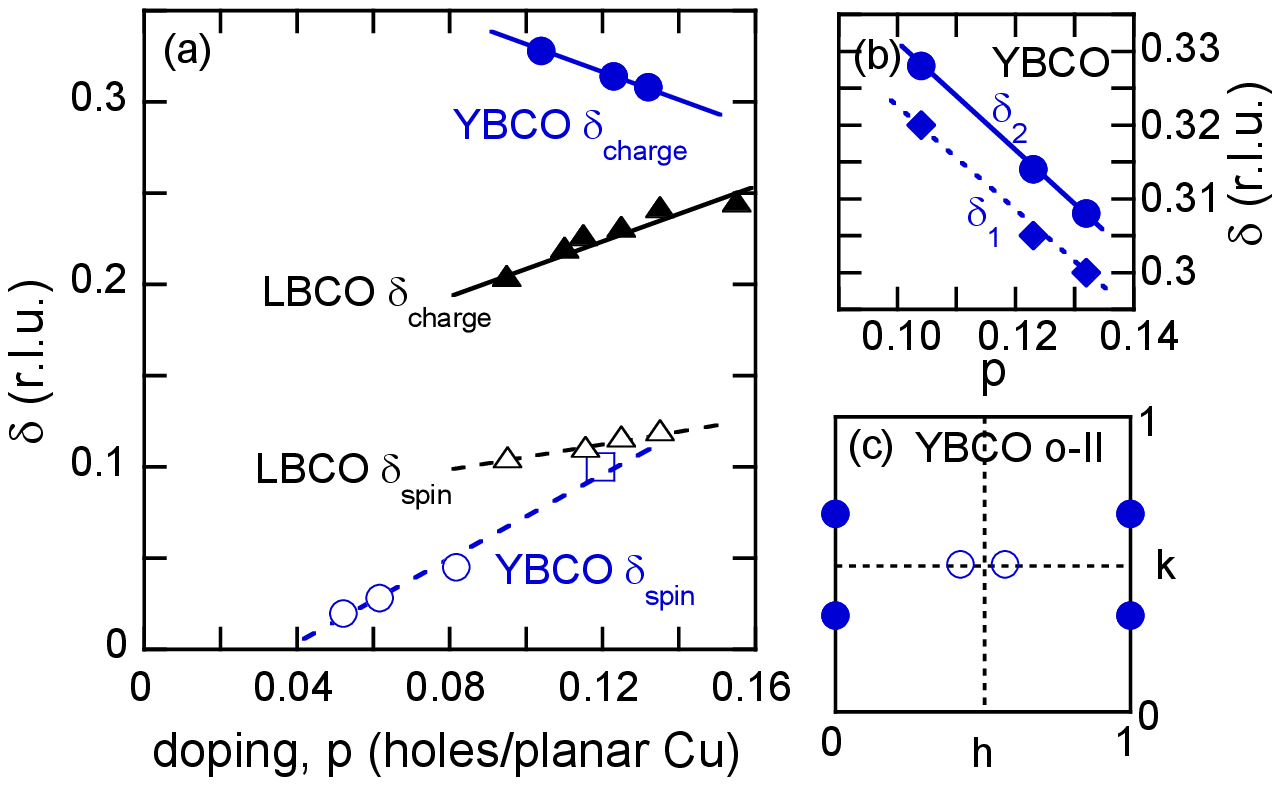}
\end{center}
\caption{(a) Spin and charge incommensurability versus doping for
YBCO and La$_{2-x}$Ba$_{x}$CuO$_{4}$ (LBCO).
The spin incommensurability of both YBCO \cite{Haug} and La$_{2-x}$Ba$_{x}$CuO$_{4}$ \cite{Hucker}
increases with doping.  In LBCO the spin and charge incommensurability are simply related: $\delta_c\approx2\delta_s$. In YBCO, the spin and charge incommensurability have opposite trends with doping.
(b) The charge incommensurability in YBCO plotted on an expanded scale.
(c) In YBCO o-II, the dominant wavevectors of the spin ($\boldsymbol{\delta}_{\mathrm{spin}}$) and charge ($\boldsymbol{\delta}_{\mathrm{charge}}$) modulations are along different directions: the {\bf a}- and {\bf b}- axes respectively.}
\label{fig:fig4}
\end{figure}

An important issue in the cuprates is the relationship of the spin and charge correlations \cite{Ghiringhelli,Chang,Hucker,Tranquada}, 
where the underlying antiferromagnetism (AF) and charge density have modulations characterised by 
wavevectors $\boldsymbol{\delta}_{\mathrm{spin}}$ and $\boldsymbol{\delta}_{\mathrm{charge}}$ respectively. In a simple stripe picture of
 intertwined spin and charge correlations \cite{Tranquada},  these yield spin and charge peaks at positions $\tau_{\mathrm{AF}} \pm \boldsymbol{\delta}_{\mathrm{spin}}$ and
$\tau_{\mathrm{lattice}} \pm \boldsymbol{\delta}_{\mathrm{charge}}$, where $\boldsymbol{\delta}_{\mathrm{charge}}= 2\boldsymbol{\delta}_{\mathrm{spin}}$.
This simple relationship appears to describe observations
in La$_{2-x}$Ba$_{x}$CuO$_4$ (see Fig.~\ref{fig:fig4}) and
La$_{1.6-x}$Nd$_{0.4}$Sr$_{x}$CuO$_4$ \cite{Hucker,Tranquada}.
In YBCO, the low-frequency spin fluctuations are anisotropic \cite{Stock,Hinkov}, with the strongest
 response for $\boldsymbol{\delta}$ along $\mathbf{a}^{*}$.  Indeed, lightly doped
YBa$_2$Cu$_3$O$_{y}$ shows magnetic order \cite{Haug} with
$\boldsymbol{\delta}$ along $\mathbf{a}^{*}$. Thus in YBCO (see Fig.~\ref{fig:fig4}), not only 
are $\boldsymbol{\delta}_{\mathrm{spin}}$ and $\boldsymbol{\delta}_{\mathrm{charge}}$ in different directions, but they show different trends and
$| \boldsymbol{\delta}_{\mathrm{charge}}| \neq 2 |\boldsymbol{\delta}_{\mathrm{spin}} |$.
These differences suggest that $\boldsymbol{\delta}_{\mathrm{spin}}$ and
$\boldsymbol{\delta}_{\mathrm{charge}}$ have different origins, e.g.~they may be
determined by different Fermi surface nesting vectors.

We may also consider $T_{\mathrm{CDW}}$ as a function of doping (Table 1), although we stress the difficulty in determining unambiguously the onset of a small signal from a large background.  In Fig. 3 (c), it is clear that for o-II, the CDW appears at a higher temperature, indicating that as the hole doping is reduced, $T_{\mathrm{CDW}}$ increases.  This pattern coincides roughly with the onset of the polar Kerr effect \cite{Xia} and the inflection point $T_H$ below which the Hall coefficient begins to fall towards negative values \cite{LeBoeufPRB}, as suggested previously \cite{Chang}.  This indicates that all these measurements are sensitive to the same lattice symmetry breaking process.

This places our observations in context with other low-field measurements.
 However, originally, the existence of CDW order was indicated by high-field NMR
work on o-II \cite{Wu}.
This was revealed by a splitting, at high magnetic field and low temperature,
of the NMR lines of the Cu2F sites, which lie in the CuO$_2$ planes next to a
filled CuO chain. This splitting was most simply explained by invoking a 1-{\bf q} CDW
 along the {\bf a} direction with $ \delta=0.25$ \cite{Wufootnote}, which could give a Cu2F
 splitting, while having no effect on the  Cu2E sites next to empty chains. However,
 very recent ultrasonic data \cite{LeBoeufNP} indicate that under similar high-field
 conditions o-II is actually 2-{\bf q}.  This is inferred from the sharp knee in
 velocity for the $c_{66}$ shear mode seen at 18 T, which can only couple to a CDW
 having components of propagation and displacement in both directions in the $\bm{a}$-$\bm{b}$ plane.
 Our present \emph{zero-field} results show that o-II exhibits a dominant CDW modulation vector different
 in magnitude and direction from that inferred from the NMR data, but similar to that seen by diffraction
at other dopings \cite{Ghiringhelli,Chang,Achkar}. Assuming that the CDW is responsible for the FS reconstruction, 
the weak doping-dependence of QO frequencies \cite{Sebastian,Vignolle} also suggests that o-II develops CDW order similar to
adjacent dopings. We are ineluctably driven to seek a new explanation for the NMR data consistent
with these other measurements.

Firstly, we note that the onset fields and temperatures given by NMR~\cite{Wu}
 and ultrasonics~\cite{LeBoeufNP} differ from those given by
X-ray diffraction~\cite{Chang,Ghiringhelli,Achkar}. As discussed extensively
elsewhere \cite{Chang,Ghiringhelli}, NMR (and ultrasonics) are relatively low-frequency
techniques, and it seems likely that quasi-static CDWs become visible to X-rays before
they become slow enough to have strong effects on NMR and ultrasonics.  It is possible
that the anomaly seen at 18 T with ultrasound \cite{LeBoeufNP} is the locking of the
CDW order that we observe into $\delta \rightarrow 1/3$, accompanied by a freezing of the CDW.

A 1-{\bf q} \emph{incommensurate} sinusoidal charge density wave along {\bf b} would
give a bimodal distribution of NMR/NQR frequencies at a given Cu site, with van Hove peaks
at the extremal values, while a 1-{\bf q} \emph{commensurate} $\delta= 1/3$ modulation can
 give two peaks of unequal weight. If barely-resolved, either could mimic a simple splitting.
In contrast, a 2-{\bf q} incommensurate pattern, with the two components identical, would give
a single central van Hove peak. However, a 2-{\bf q}
pattern with one component having rather weaker effects than the other, could again give
a bimodal distribution, similar to that observed for the Cu2F sites. A question for any modulation
along {\bf b} is to explain the much smaller effects of the CDW on the Cu2E sites.
This could arise if the CDW caused larger displacements of the atoms (e.g. oxygen O3) near
 the Cu2F sites as is allowed by symmetry in the full o-II unit cell. Alternatively, the
 stronger effects of the CDW at the Cu2F sites may be related to their proximity to the conducting
 charge-reservoirs represented by the well-ordered Cu-O chains.
We put forward this model in an attempt to reconcile apparently conflicting results from different
 methods.
At present diffraction data cannot be taken at sufficiently high
 steady magnetic fields to provide
structural data to complement the spectroscopic data from NMR, the thermodynamic data
from ultrasonics and the Fermi surface data from QO.


In summary, we have detected CDW ordering in ortho-II YBCO in zero magnetic
field.  The major component of the ordering has been
found to have $\mathbf{q}_{\mathrm{CDW}}$ with an in-plane component along the {\bf b} (chain)
direction. This contrasts with nearby higher dopings having less perfect Cu-O chain
 order, where 2-$\mathbf{q}$ structures are observed in which the two modulations have similar amplitudes.
The incommensurability and $T$-dependence of the CDW order are very similar
 to those previously reported in o-VIII and -III YBCO, but there is a clear trend to a
 larger $q_{\mathrm{CDW}}$ at lower doping;
this suggests a band-structure influence on  $q_{\mathrm{CDW}}$.
 Our observation in o-II YBCO of a dominant charge modulation along {\bf b} strongly suggests that a simple spin/charge stripe picture may not be appropriate, since the incipient spin correlations have a wavevector along {\bf a}.  The independent values
 of the spin and charge correlation {\bf q}-vectors over a range of dopings indicates that these have
 different origins in YBa$_2$Cu$_3$O$_{y}$.
For YBCO o-II, the dominant modulation direction in zero field and its incommensurability
 are completely different from those inferred from high-field NMR data. We propose
 an alternative explanation of these data.  

{\it Note Added.} A recent preprint \cite{Blanco}
 reports a soft x-ray study of a similar o-II YBCO sample. 
The authors find a modulation along $\mathbf{b}$, together with a weaker modulation along $\mathbf{a}$. 
A weak modulation along $\mathbf{a}$ has been revealed by further hard x-ray experiments by ourselves at the more 
intense beamline P07 on PETRA-III at DESY. The value of $\delta_1$ given by our measurements is in 
Table~I and Fig. 4.

\begin{acknowledgments}
This work was supported by the EPSRC (grant numbers EP/G027161/1, EP/J015423/1 \& EP/J016977/1), the Wolfson Foundation, 
the Royal Society, the Office of Basic Energy Sciences, U.S. Department of Energy, under Contract No. DE-AC02-98CH10886,
 the Danish Agency for Science, Technology and Innovation under DANSCATT
and the Swiss National Science Foundation through NCCR-MaNEP and grant
number PZ00P2\_142434. 
We thank M.W.\,Long, C.\,Proust, B.\,Vignolle and D.\,LeBoeuf for discussions.
\end{acknowledgments}

\end{document}